\documentclass[conference]{IEEEtran}
\IEEEoverridecommandlockouts
\usepackage{fancyhdr}
\usepackage{cite}
\usepackage{amsmath,amssymb,amsfonts}
\usepackage{algorithmic}
\usepackage{graphicx}
\usepackage{textcomp}
\usepackage{xcolor}
\def\BibTeX{{\rm B\kern-.05em{\sc i\kern-.025em b}\kern-.08em
    T\kern-.1667em\lower.7ex\hbox{E}\kern-.125emX}}

\usepackage{dblfloatfix}

\usepackage{tikz}
\usepackage{setspace}
\newcommand{\subparagraph}{}
\usepackage{titlesec}
\usepackage{tablefootnote}

\usepackage[caption=false,font=footnotesize]{subfig}

\titlespacing\section{1pt}{3pt plus 4pt minus 2pt}{1pt plus 2pt minus 2pt}
\titlespacing\subsection{1pt}{3pt plus 4pt minus 2pt}{1pt plus 2pt minus 2pt}
\titlespacing\subsubsection{1pt}{3pt plus 4pt minus 2pt}{1pt plus 2pt minus 2pt}

\newcommand*\circled[1]{\tikz[baseline=(char.base)]{
            \node[shape=circle,fill,inner sep=0.7pt] (char) {\textcolor{white}{#1}};}}
            
\usepackage[yyyymmdd,hhmmss]{datetime}

\usepackage{multirow}

\newcommand{\lois}[1]{\textcolor{black}{#1}}

\newcommand{\lon}[1]{\textcolor{black}{#1}}

\newcommand{\joruge}[1]{\textcolor{black}{#1}}
\newcommand{\jorrge}[1]{\textcolor{black}{#1}}

\newcommand{\ruth}[1]{\textcolor{black}{#1}}
\newcommand{\ruthy}[1]{\textcolor{black}{#1}}

\newcommand{\jorge}[1]{\textcolor{black}{#1}}

\newcommand{\maarten}[1]{\textcolor{black}{#1}} 
\newcommand{\onur}[1]{\textcolor{black}{#1}}

\newcommand{\cready}[1]{\textcolor{black}{#1}}
\newcommand{\creadyy}[1]{\textcolor{black}{#1}}

\begin{document}

\bstctlcite{IEEEexample:BSTcontrol}
\title{\textbf{Optically Connected Memory \\for Disaggregated Data Centers}%
\thanks{\jorrge{This work is supported by the LPS Advanced Computing Systems (ACS) Research Program  (contract} HD TAT DO 7 (HT 15-1158)), the Department of Energy (DOE) Small Business Innovation Research (SBIR) ASCR Program (contract DE-SC0017182), the Sao Paulo Research Foundation (FAPESP) (fellowships 2013/08293-7 and 2014/01642-9), CAPES (fellowships 2013/08293-7 and 88882.329108/2019-01), and CNPq (fellowships 438445/2018-0,  309794/2017-0 and 142016/2020-9).
}
}
\author{
\IEEEauthorblockN{Jorge~Gonzalez\IEEEauthorrefmark{1}\enspace
  Alexander~Gazman\IEEEauthorrefmark{3}\enspace
  Maarten~Hattink\IEEEauthorrefmark{3}\enspace
  Mauricio~G.~Palma\IEEEauthorrefmark{1}\enspace
  Meisam~Bahadori\IEEEauthorrefmark{4}\enspace
  \\Ruth~Rubio-Noriega\IEEEauthorrefmark{5}\enspace
  Lois~Orosa\IEEEauthorrefmark{7}\enspace
  Madeleine~Glick\IEEEauthorrefmark{3}\enspace
  Onur~Mutlu\IEEEauthorrefmark{7}{}\enspace
  Keren~Bergman\IEEEauthorrefmark{3}\enspace
  Rodolfo~Azevedo\IEEEauthorrefmark{1}}
\IEEEauthorblockA{\\
\IEEEauthorrefmark{1}{University of Campinas}
\quad\IEEEauthorrefmark{3}{Columbia University}
\quad\IEEEauthorrefmark{4}{Nokia}
\quad\IEEEauthorrefmark{5}{INICTEL-UNI}
\quad\IEEEauthorrefmark{7}{ETH Z{\"u}rich}}
}

%
\definecolor{MidnightBlue}{rgb}{0.1, 0.1, 0.44}
\newcommand{\versionnum}[0]{4.0~\today~@ \currenttime~(Zurich time)} 
\newif\ifcameraready
\camerareadytrue

\maketitle

\fancyhead{}
\ifcameraready
 \thispagestyle{plain}
 \pagestyle{plain}
\else
 \fancyhead[C]{\textcolor{MidnightBlue}{\emph{Version \versionnum~---~}}}
 \fancypagestyle{firststyle}
 {
   \fancyhead[C]{\textcolor{MidnightBlue}{\emph{Version \versionnum~---~}}}
   \fancyfoot[C]{\thepage}
 }
 \thispagestyle{firststyle}
 \pagestyle{firststyle}
\fi
\begin{abstract}
Recent advances in integrated photonics enable the implementation of reconfigurable, high-bandwidth, and low energy-per-bit interconnects in next-generation data centers. We propose and evaluate an Optically Connected Memory (\textbf{OCM}) architecture that disaggregates the main memory from the computation nodes in data centers. OCM is based on micro-ring resonators (MRRs), and it does not require any modification to the DRAM memory modules. We calculate energy consumption from real photonic devices and integrate them into a system simulator to evaluate performance. Our results show that (1) OCM is capable of interconnecting four DDR4 memory {channels} to a computing node using two fibers with $1.07$ pJ energy-per-bit consumption and (2) OCM performs up to \cready{5.5}$\times$ faster than a disaggregated memory with 40G PCIe NIC connectors to computing nodes.
\end{abstract}

\begin{IEEEkeywords}
disaggregated computing, disaggregated \cready{memory}, photonics, data-centers, \cready{DRAM}, \creadyy{memory systems}
\end{IEEEkeywords}

\section{Introduction}\label{sec:introduction}

\begin{figure}[h]
  \centering
  \includegraphics[width=0.99\linewidth]{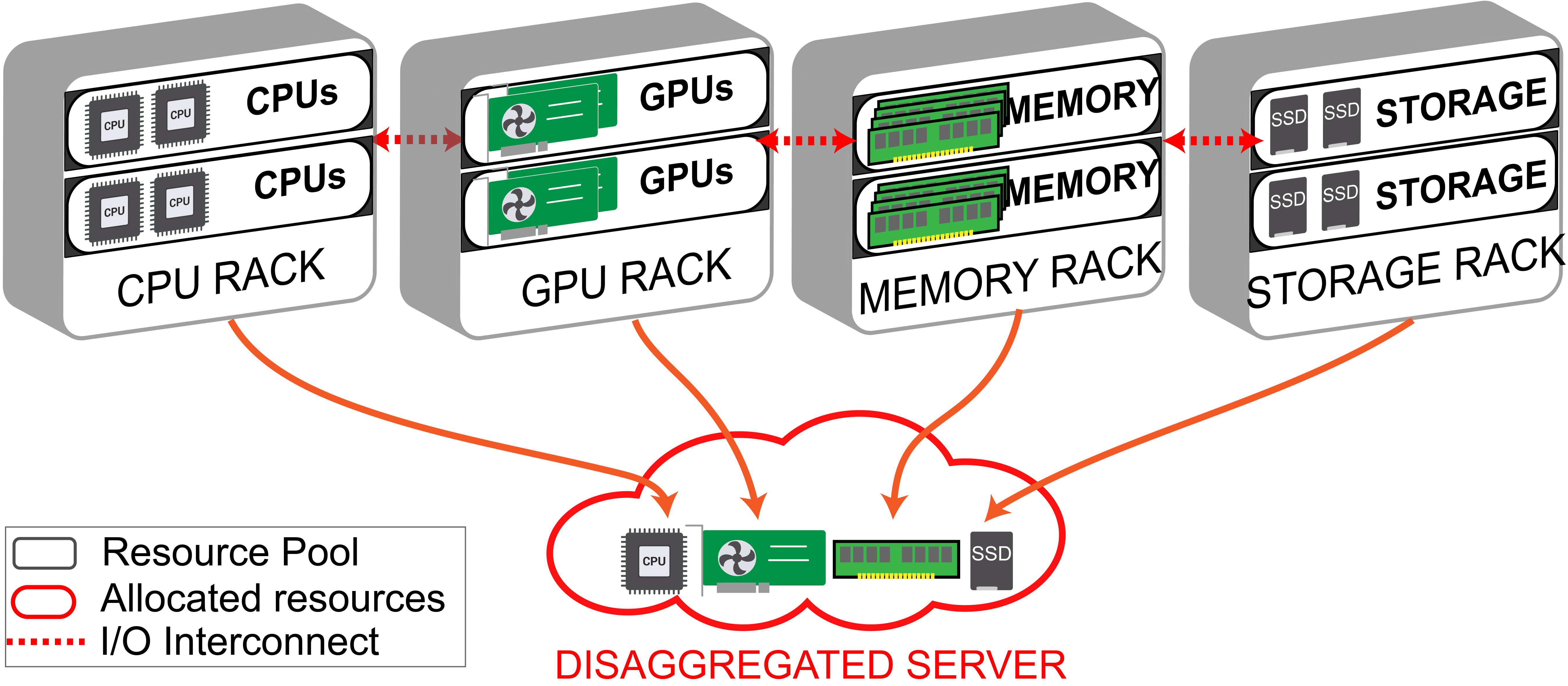}
  \caption{Disaggregation concept for data centers.}
  \label{fig:disaggregated_pine}
\end{figure}

\IEEEPARstart{S}{}caling and maintaining conventional memory systems in modern data centers is challenging for three fundamental reasons. First, the dynamic memory capacity demand is difficult to predict in the short, medium, and long term. As a result, memory capacity is usually over-provisioned\creadyy{~\cite{Luo2020,panwar2019quantifying,Chen2018,reiss2012towards,Di2012}}, which wastes resources and energy. Second, workloads are limited to using the memory available in the local server (even though other servers might have unused memory), which could cause memory-intensive workloads to slow down. Third, memory maintenance might cause availability issues\creadyy{~\cite{Meza2015}}; in case a memory module fails, all running applications on the node \cready{may} have to be interrupted to replace the faulty module. A promising solution to overcome these issues is to disaggregate the main memory from the computing cores{~\cite{lim2009disaggregated}}. As depicted in Figure \ref{fig:disaggregated_pine}, the key idea is to organize \cready{and cluster} the memory resources such that they are individually addressable and accessible from any processor in the data center \cite{bergman2018pine}. Memory disaggregation {provides} flexibility in memory allocation, improved utilization of the memory resources, lower maintenance costs, and lower energy consumption \onur{in} the data center \cite{papaioannou2016benefits}.

\ruth{Disaggregating memory and processors remains a challenge, although the disaggregation of some resources (e.g., storage) is common in production data centers\creadyy{~\cite{legtchenko2017understanding}}.} Electrical interconnections in rack-distances do not fulfill the low latency and \onur{high bandwidth} requirements of modern \cready{DRAM} modules. The primary limitation of an electrical interconnect is that it constrains the memory bus to onboard distance~\cite{endofmoorelaw} \ruth{because the \creadyy{electrical wire's signal integrity loss} increases at higher frequencies.} This loss dramatically reduces the Signal-to-Noise Ratio (SNR) when distances are large. An optical interconnect is more appealing than an electrical interconnect for memory disaggregation due to three properties: its (1) high bandwidth density significantly reduces the number of IO lanes, (2) power consumption and crosstalk do {\emph{not}} increase with distance, and (3) propagation loss is low. {Silicon Photonic (SiP)} devices are likely suitable for disaggregation, delivering $\geq$ Gbps range bandwidth, as well as efficient and versatile switching \cite{}.

\ruthy{The \textbf{goal} of this work is to pave the way for designing high-performance \textit{optical memory channels} (i.e., the optical equivalent of an electrical memory channel) that enable main memory disaggregation in data centers.}
\cready{Our work \creadyy{provides} an optical link design for DDR DRAM memory disaggregation, and it defines its physical characteristics, i.e., i) number of \creadyy{Micro-Ring Resonator (MRR)} devices, ii) bandwidth per wavelength, iii) energy-per-bit, and iv) area. \creadyy{We evaluate the performance (see Section \ref{sec:sysleveleval}) and energy consumption (see Section \ref{sec:sip_section}) of a system with  \creadyy{disaggregated commodity} DDR DRAM modules.}} 

We make three key contributions: (1) we propose the Optically Connected Memory (OCM) architecture for memory disaggregation in data centers based on state-of-the-art photonic devices, (2) we perform the first evaluation of the energy-per-bit consumption of a SiP link using the bandwidth requirements of current DDR \cready{DRAM} standards, and (3) we model and evaluate OCM in a system-level simulator \joruge{and show that it performs up to 5.5x faster than a 40G NIC-based disaggregated memory.}

\section{Motivation}

\jorge{Photonics is very appealing for memory disaggregation because: (1) the integration (monolithic and hybrid) between electronics and optics has already been demonstrated~\cite{absil2015imec}, which \cready{allows} the design and fabrication of \creadyy{highly-integrated} and complex optical subsystems on a chip, 
\cready{and (2)} optical links offer better scaling in terms of bandwidth, energy, and IO compared to electrical links; e.g., \cready{optical switches (o-SW)} show better port count scaling~\cite{sato2018realization}).} 

\lois{New electrical interfaces, such as GenZ, CCIX, and OpenCAPI, can disaggregate a wide range of resources (e.g., memory, accelerators)~\cite{benton2017ccix}.} 
\lois{Optical devices can enable scalable rack-distance, and energy-efficient interconnects for these new interfaces, as demonstrated by a previous work that disaggregates the PCIe interface with silicon photonics~\cite{zhu2019flexible}.}
\jorrge{\lois{Our} OCM \lois{proposal} extends the memory interface with optical devices and does not require substantial modifications to it, e.g., the memory controllers remain on the compute nodes.} 
%
%
%



\begin{figure}[!h]
  \centering
  \includegraphics[width=\linewidth]{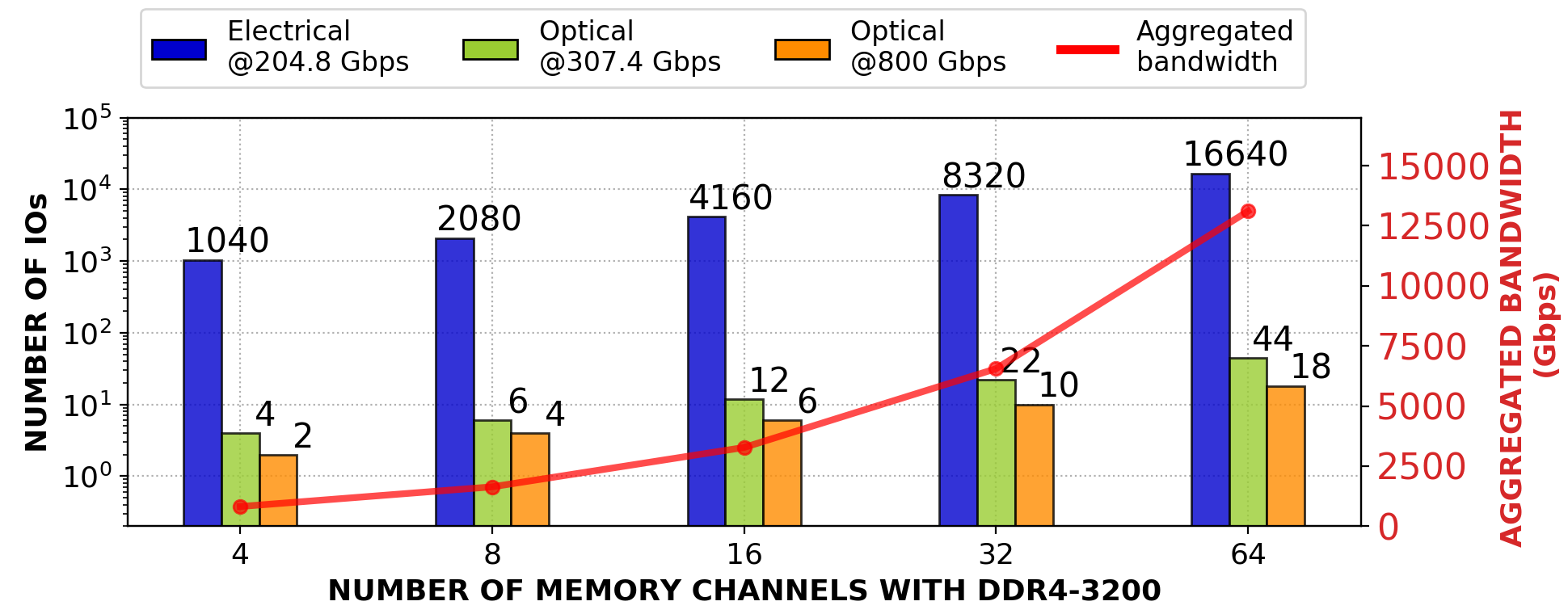}
  \caption{Required electrical and optical IO counts (lower is better) for sustaining different \onur{amounts of} aggregated \onur{bandwidth}.}
  \label{fig:pinesandpins}
\end{figure}

\jorge{Figure \ref{fig:pinesandpins} shows the IO requirements in the memory controller for electrical \cite{marino2018architectural}, and optical interconnects to achieve a specific aggregated bandwidth. We define IO as the number of required electrical wires or optical fibers in the interconnects. We use, for both electrical and optical interconnects, 260-pin DDR4-3200 DRAM modules with 204.8 Gbps maximum bandwidth per memory channel. We make two observations. First, the required number of optical IOs (left y-axis) is up to three orders of magnitude smaller than the electrical IOs because an optical fiber can contain many \emph{virtual channels} using Wavelength Division Multiplexing (WDM) \cite{57798,bahadori2016comprehensive}. Second, a single optical IO achieves up to 800 Gbps based on our evaluation, requiring 2 IOs for bidirectional communication (see Section \ref{sec:sip_section}). An optical architecture could reach the required throughput for a 4 memory channel system using only 2 IOs (two fibers) and for a 32-channel system with only 10 IOs.} 

\section{OCM: Optically Connected Memory}\label{arch_section}

To overcome the electrical limitations \cready{that can potentially impede} memory disaggregation, we introduce an OCM that does not require modifications in the \cready{commonly-used} DDR \cready{DRAM} protocol. OCM places commodity DRAM Dual Inline Memory Modules (DIMMs) at rack-distance from the processor, and it sustains multiple memory channels by using different wavelengths for data transmission. OCM uses conventional DIMMs and memory controllers, electro-optical devices, and optical fibers to connect the computing cores to the memory modules. 
Our work explores the idea of direct point-to-point optical interconnects for memory disaggregation and extends \cready{prior works}  \cite{brunina4,anderson2018reconfigurable}, to reduce \cready{the} latency overhead caused by additional protocols such as remote direct memory access (RDMA) and PCIe \cite{Zervas:18}.
OCM is versatile and scales with the increasing number of wavelengths per memory channel expected from future photonic systems~\cite{Glick:18}.

\begin{figure*}[t]
  \centering
  \includegraphics[width=1.05\textwidth]{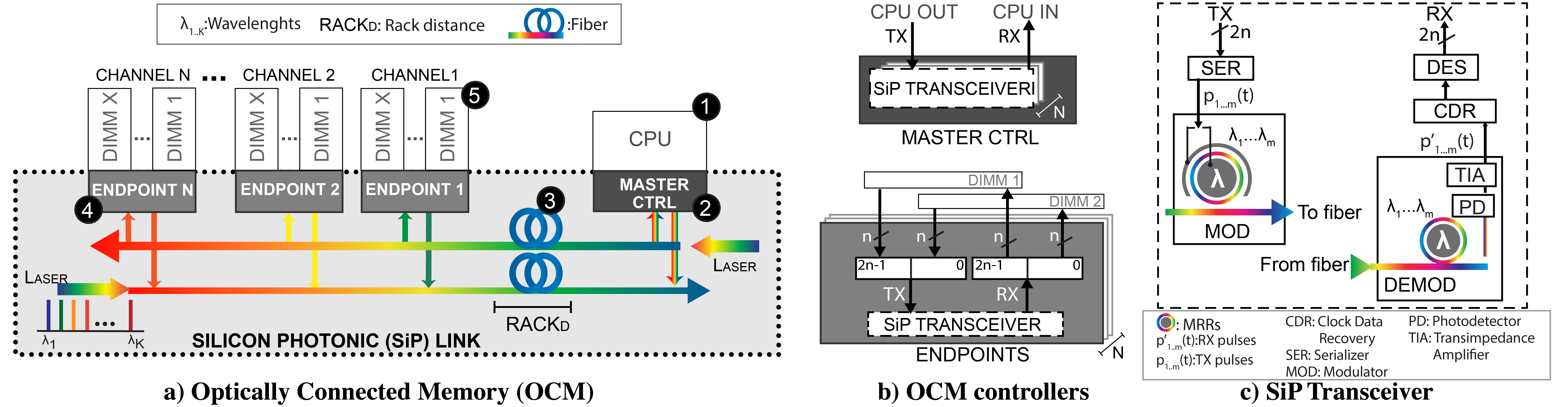}
  \caption{Optically Connected Memory organization: optical memory channels for disaggregation of the main memory system.}
  \label{fig:ocm}
\end{figure*}

\subsection{Architecture Overview}\label{sec:archoverview}

Figure~\ref{fig:ocm}a shows the main components of the OCM architecture configured with state-of-the-art photonic devices and DDR memories. \ruth{OCM uses N optical memory channels, each one consisting of X memory modules (DIMM 1 to X) operating in lockstep.} OCM uses two key mechanisms to take advantage of the high aggregated bandwidth of the optical domain while minimizing the \cready{electrical-optical-electrical} conversion latency overhead. First, it implements an optical memory channel with multiple wavelengths that can support multiple DIMMs in a memory channel. Second, it achieves high throughput by increasing the cache line size and splitting it across all the DIMMs in a memory channel. For example, if OCM splits a single cache line between two DIMMs, it halves the bus latency (i.e., data burst duration $tBL$), compared to a conventional DDR memory.

In our evaluation (Section~\ref{sec:eval}), we use two DDR channels operating in lockstep to get a cache line of 128 bytes with similar latency as a cache line of 64 bytes in a single DDR channel (Section \ref{sec:timingmodel}). \ruth{OCM benefits from the use of a wide $Xn$-bit interface, where $X$ is the number of DIMMs, and $n$ is the width in bits of a DIMM bus.} OCM transfers depend on the serialization capabilities of the SiP transceiver.  The serialization/deserialization latency increases with the number of DIMMs in lockstep. Notice that, a commercial SERDES \cready{link} (e.g., ~\cite{hadidi2017demystifying}) supports serialization up to 256B (i.e., four 64B cache lines). As shown in Figure \ref{fig:ocm}a, on the CPU side, there is a Master controller, and on the memory side, there are N Endpoint controllers that respond to CPU requests. Both controllers have a structure called SiP Transceiver, and Figure \ref{fig:ocm}b shows a difference in the organization of the SiP transceivers per controller. Figure \ref{fig:ocm}c shows the SiP transceivers present in the Transmitter (TX) and Receiver (RX) lanes \cready{in} both Master and Endpoint controllers. A TX lane consists of a serializer (SER) and Modulator (MOD) for transmitting data. An RX lane contains a Demodulator (DEMOD), a Clock and Data Recovery (CDR) block, and a Deserializer (DES) for receiving data. \ruth{Both TX and RX lanes connect with a $Xn$-bit (e.g., $X$=2 and $n$=64 in our evaluation) bus to the Endpoint controller, which forms the bridge between the lanes and the DRAM module.}

\begin{figure*}[!b]
  \centering
  \includegraphics[width=0.65\linewidth]{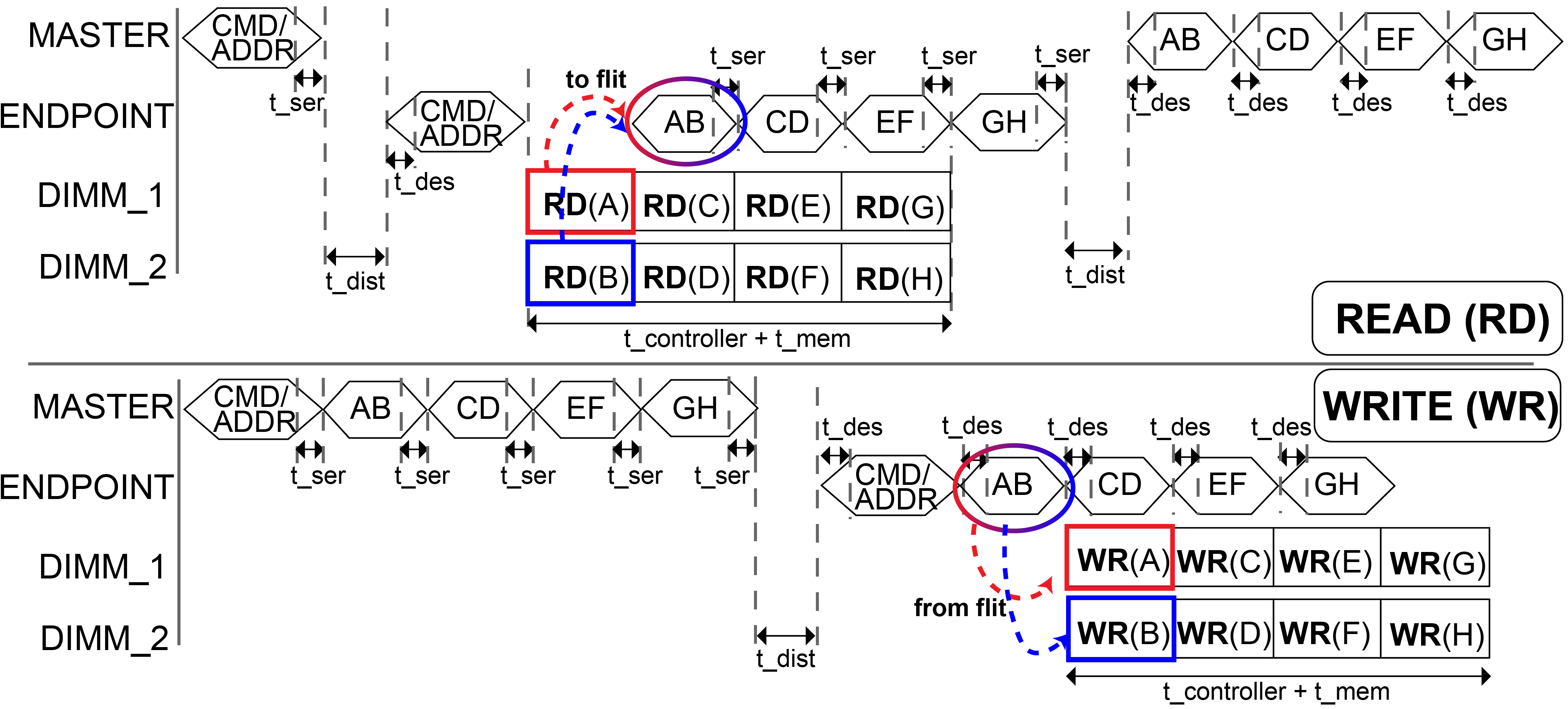}
  \caption{OCM timing diagram for Read (top) and Write (bottom) \creadyy{requests.}}
  \label{fig:time}
\end{figure*}

\subsection{Timing \creadyy{Model}}\label{sec:timingmodel}
\jorge{OCM transfers a cache line as a serialized packet composed of smaller units called \textit{flits}, \cready{whose} number depends on the serialization capabilities of the SiP transceiver. Figure \ref{fig:time} presents the timing diagram of the OCM Read (RD) and Write (WR) operations. For reference, a conventional DDR \cready{DRAM} memory channel uses 64B cache lines; a data bus transfers each line as 8B data blocks in 8 consecutive cycles, and the 1B Command (CMD) and 3B Address (ADDR) use separate dedicated buses.}
\jorge{In OCM, as depicted in Figure \ref{fig:time}, the cache line is transferred in AB-GH flits. We show OCM timing with a $flit$ size that doubles the width of the memory channel data bus, and is the reason for dividing the cache line between DIMMs 1 and 2 to perform parallel access and decrease latency. OCM splits a single cache line between two DIMMs, which halves the bus latency (i.e., $tBL$ \cite{jedecddr4}), compared to conventional DDR \cready{DRAM} memory. }


\jorge{For the RD operation, data A and B are read from different DIMMs to compose a flit (AB). Flit AB serialization and transmission occur after the Master controller receives the CMD/ADDR flit. For the WR operation, the Master controller sends the flit containing data blocks AB immediately after the CMD/ADDR flit. After Endpoint deserialization, DIMM 1 stores A, and DIMM 2 stores B. For example, OCM with a commercial \creadyy{Hybrid Memory Cube (HMC)} serializer~\cready{\cite{hadidi2017demystifying}} and 128B cache line size, transfers 2 $\times$ (4 $\times$ 16B of data) with 1 $\times$ 4B CMD/ADDR initiator message (or \textit{extra flit}).}


Compared to conventional electrical DDR memory, OCM adds serialization and optical {packet transport latency} to the overall memory access time (see Section~\ref{sec:eval}).
\jorrge{The DIMM interface can support the latency overhead that is imposed by our optical layer integration. In our evaluation, we consider both optimistic and worst-case scenarios. \cready{Past experimental} works \cite{anderson2018reconfigurable} \cready{show that} the overhead is low in the order of a few nanoseconds, requiring no modification to the memory controller. However, if there is high latency imposed by the optical layer, the signaling interface from the memory controller needs to be adapted.}
Equation \ref{time_total} shows the OCM latency model $T_{lat}$, which is defined as the \onur{sum} of the DIMM controller latency $T_{contr}$, DIMM WR/RD latency \maarten{ $T_{mem(A|B)}$ (latency is equal for both DIMMs)}, serialization/deserialization latency $T_{serdes}$, modulation/demodulation {latencies $T_{mod}$ and $T_{demod}$}, distance propagation latency penalty $T_{dist}$, and  system initialization time (e.g., Clock Data Recovery (CDR) {latency}, modulator resonance locking~\cite{padmaraju2013wavelength}) $T_{setup}$.
\begin{equation}
\label{time_total}
\begin{aligned}
T_{lat}(t)=&T_{setup}+T_{contr}+T_{mem(A|B)}(t)+T_{serdes}\\
&+T_{mod}+T_{demod}+T_{dist}
\end{aligned}
\end{equation}
$T_{setup}$ equals zero because it has no impact on the system once it is configured~\cite{anderson2018reconfigurable}. In the optical and millimeter wavelength bands, \ruthy{${T_{mod}}$ and {$T_{demod}$ are} \cready{in} the order of $ps$\cite{bahadori2016comprehensive}}, due to {the} small footprint {of ring modulators} (tens of micrometers) and \onur{the} high dielectric constant of silicon.

\subsection{Operation}\label{sec:ocmoperation}

Figure \onur{\ref{fig:ocm}a} illustrates the five stages of a \cready{memory} transaction.

\noindent\rule{\columnwidth}{0.7pt}
\noindent\textbf{Stage \circled{1}}: the processor generates a Read/Write (RD/WR) memory request. In the photonic domain, a laser source generates light in $\lambda_{1,2,\dots,K}$ wavelengths simultaneously~\cite{bahadori2017energy}.\\
\noindent\rule{\columnwidth}{0.7pt}
\textbf{Stage \circled{2}}: the data from the processor is serialized (SER) \maarten{onto} the Master Controller's TX lane, and the generated electrical pulses $p_{1,2,...,m}(t)$ \maarten{drive} the cascaded array of \cready{Micro-Ring Resonators (MRRs)}  for modulation (MOD), represented as rainbow rings. We use non-return-to-zero on-off keying (NRZ-OOK) that represents logical ones and zeros imprinted on the envelope of light~\ruthy{\cite{bahadori2016comprehensive}}.\\ 
\noindent\rule{\columnwidth}{0.7pt}
\textbf{Stage \circled{3}:} the optical signal is transmitted through an optical fiber. At the end of the fiber, the combined optical WDM channels are coupled into an optical receiver.\\
\noindent\rule{\columnwidth}{0.7pt}
\textbf{Stage \circled{4}:} \maarten{first, \onur{in} the RX lane of an Endpoint, the WDM Demodulator (DEMOD) demultiplexes the optical wavelengths using $m$ MRRs. Each MRR works as an optical band-pass filter to select a single optical channel from $\lambda_{1,2, ... m}$. Second, these separated channels are then fed to DEMOD's integrated photo-detectors (PD) followed by transimpedance amplifiers (TIA). Together the PD and TIA convert and amplify the optical signal to electrical pulses $p'_{1,2,...,m}(t)$ suitable for sampling. Third, the data is sampled, deserialized (DES), and sent to the memory controller.}\\ 
\noindent\rule{\columnwidth}{0.7pt}
\textbf{Stage \circled{5}:} the processor \cready{accesses} memory with the DDR protocol using a RD or WR command and a memory address. For a RD command, the Endpoint TX transmits to the processor a $cacheline$ with the wavelengths $\lambda_{1,...,m}$ (similar to Stages 1 to 4). For a WR command, the data received from the processor is stored in memory.\\

\subsection{Enabling Reconfigurability}\label{sec:reconfigurability}

\joruge{OCM supports reconfigurability by placing an \cready{o-SW} between the Endpoints and the Master controller, similar to previous \cready{work}~\cite{anderson2018reconfigurable}.}
OCM uses optical switching to connect or disconnect a master controller from an endpoint. Switching can happen (1) in the setup phase, which is the first time that the system is connected before starting execution, or (2) before executing a workload, to adapt the amount of assigned memory to the requirements of the workload.

\begin{figure}[h]
  \centering
  \includegraphics[width=0.4\textwidth]{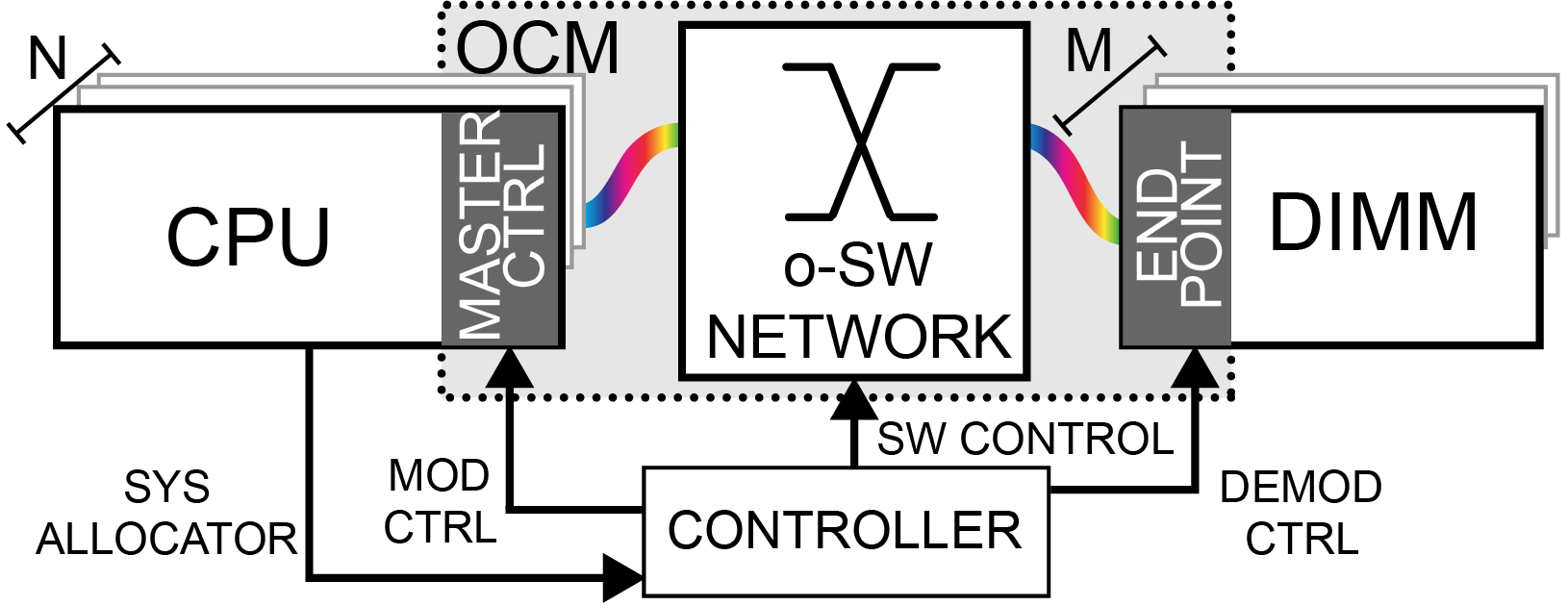}
  \caption{\joruge{Reconfigurable OCM with optical switches (o-SW).}}
  \label{fig:ocmsw}
\end{figure}

\joruge{As depicted in Figure \ref{fig:ocmsw}, an optical switch has multiple ports, through which a set of N processors can be connected to a configurable set of M OCMs, where N and M depend on the aggregated bandwidth of the SiP links.}
In Section \ref{sec:eval}, we evaluate OCM with a single CPU, \cready{and assume} that the setup phase is already completed.
\subsection{High \cready{Aggregated Bandwidth}}\label{aggbandwidth}
OCM uses WDM {\cite{bahadori2016comprehensive,57798}} \cready{to optimize} bandwidth utilization. WDM splits data transmission into multiple colors of light (i.e., wavelengths, $\lambda$s). 

To modulate data into \cready{lightwaves}, we use Micro-Ring Resonator (MRR) electro-optical modulators, which behave as narrowband resonators that select and modulate a single wavelength. We use MRRs because they have a small \cready{hardware} footprint and low power consumption~\cite{bahadori2017energy}, and \joruge{they are tailored} to work in the communications C-band (1530-1565 nm). For more detail on photonic devices, \cready{please see} \cite{Glick:18,extremescale,bahadori2018thermal}.

OCM achieves high aggregated bandwidth by using multiple optical wavelengths $\lambda_{1,2,...,K}$  (see laser in Figure~\ref{fig:ocm}a) via WDM in a single link. \ruth{The K wavelengths are evenly distributed among the controllers, where the TX/RX lanes of a single DDR memory channel have the same number ($m$) of optical wavelengths  ($\lambda_{1,2,...,m}$, see Figure~\ref{fig:ocm}c)}. 
All wavelengths have the same bit rate $b_r$, and the aggregated bandwidth for $N$ memory channels is $BW_{aggr}=b_r \times m \times N$. Assuming that $BW_{aggr}$ is higher than the required bandwidth for a single memory channel $BW_{mc}$, then $BW_{aggr}=BW_{mc} \times N$. The total number of MRRs is $2\times2\times2\times N\times m$ because each TX or RX lane requires $m$ MRRs. OCM has two unidirectional links\cready{;} each link needs both TX and RX lanes, and these lanes are located in both Endpoint controllers and Master controllers.



\section{\lon{Evaluation}}\label{sec:eval}

To evaluate system-level performance, we implement OCM architecture in the ZSIM simulator \cite{Sanchez:2013:ZFA:2485922.2485963}. To evaluate the interconnection between processor and memory as a point-to-point SiP link, we use PhoenixSim \cite{rumley2016phoenixsim} with parameters extracted from state-of-the-art optical devices \cite{bahadori2016comprehensive,7462281,7483019}. The SiP link energy-per-bit modeling allows us to find: (1) the required number of optical wavelengths ($\lambda$), and (2) the bit rate per $\lambda$. \jorge{Table \ref{table:sip} lists OCM optical devices and their main characteristics used in our simulation model.}

\begin{figure*}[!b]
  \centering
  \includegraphics[width=0.75\linewidth]{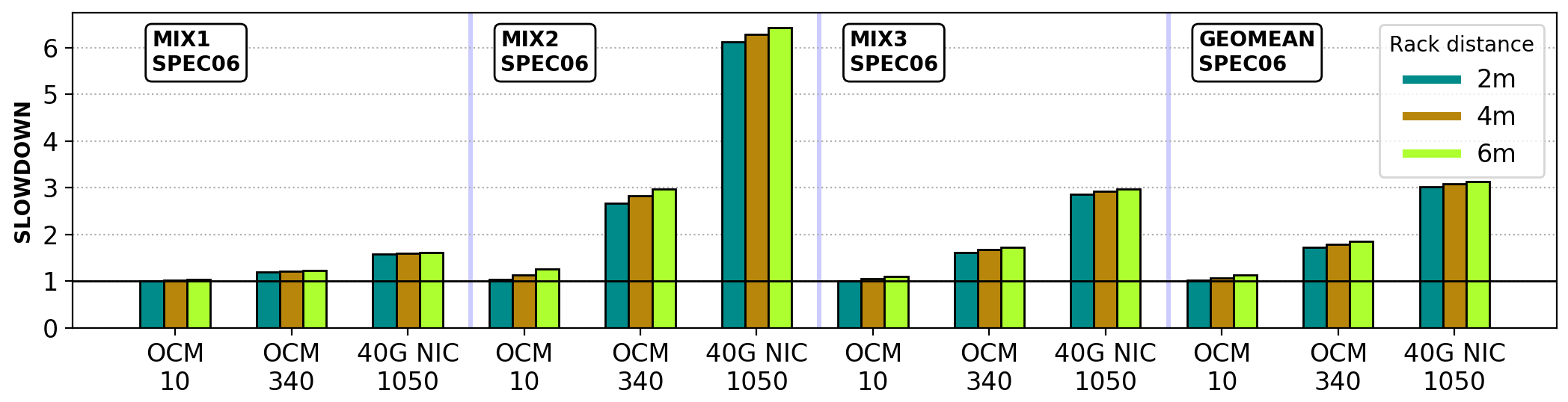}
  \caption{Slowdowns of OCM and \cready{40G NIC-based} disaggregated systems, compared to a non-disaggregated baseline with \cready{MemConf1}, for three \cready{randomly-selected} mixes of SPEC06 benchmarks (lower is better).}
  \label{fig:performance_single}
\end{figure*}

\begin{table}[h]
\scriptsize
\caption{\cready{Optical and electrical models for} OCM SiP \cready{link} devices}
\begin{tabular}{p{2cm}p{2cm}p{2.6cm}p{.5cm}}
\hline
Parameter  & Design Criteria & Details & Ref.\\
\hline 
Optical power  & 20 dBm & Max. aggregated &\\
Center wavelength        & 1.55 $\mu$m       \\
Laser & 30\% & Laser wall-plug efficiency & \cite{8478238}\\
Waveguide loss & 5 dB/cm & fabrication roughness & \cite{grillot2004size}\\
 & 0.02 dB/bend & waveguide bend loss & \\
Coupler loss & 1 dB & off-chip coupler & \cite{chen2011subwavelength}\\
Modulator   &  Q = 6500  & Ring resonator Q factor   &   ~\cite{7462281}  \\
    & ER = 10 dB & MRR extinction rate & \\
    & 65 fF & Junction capacitance  &   \\
    & -5 V & Maximum drive voltage    &   \\
   
   & 1 mW & Thermal-tuning power/ring & \cite{bahadori2018thermal}\\
Mod. mux and receiver demux   & MRR power penalties & Crosstalk model & \cite{bahadori2016comprehensive}\\
Photodetector &  1 A/W   & Current per o-power   & \cite{vivien200942} \\
\hline
Modulator driver & 28 nm & Semicond. tech. for OOK-WDM & ~\cite{7462281}   \\
SERDES power model & 28 nm & Semicond. tech. & ~\cite{7462281} \\
Digital receiver & 28 nm & Semicond. tech. for OOK-NRZ  & ~\cite{7462281}  \\
    

Element positioning & 100 $\mu$m & Modulator padding\\
\hline

\end{tabular}
\label{table:sip}
\end{table}


\joruge{Table \ref{table:zsim} shows the configuration of our baseline system (a server processor), the two DDR4 memory configurations used in our evaluation (MemConf1 and MemConf2), the latencies of an OCM disaggregated system, and the latencies of a disaggregated system using \joruge{40G PCIe NICs}. MemConf1 has 4 DDR4 memory channels as in conventional server processors, and MemConf2 has} \lois{a single DDR4 memory channel, and an in-package DRAM cache \cready{on} the processor side. The goal of the DRAM cache is \cready{to reduce} the optical disaggregation overhead \cite{Zervas:18}, which can have a significant performance impact in memory-bound applications. Our DRAM cache resembles \cready{the} Banshee DRAM cache~\cite{yu2017banshee} that tracks the contents of the DRAM cache using TLBs and page table entries, and replaces pages with a frequency-based predictor mechanism. We configure our DRAM cache to have the same operation latency as commodity DDR4 memory.}

\begin{table}[h]
\scriptsize
\tabcolsep=0.07cm
\caption{Baseline processor, memory, OCM, and NIC.}
\centering
\begin{tabular}{r|r|l}
\hline
\multirow{2}{*}{Baseline}&\textit{Processor}&3 GHz, 8 cores, 128B cache lines\\
& \textit{Cache} &32KB L1(D+I), 256KB L2, 8MB L3\\\cline{1-3}
MemConf1&\textit{Mem}&4 channels, 2 DIMMs/channel, DDR4\creadyy{-2400}\creadyy{~\cite{jedecddr4}}\\\cline{1-3}
\multirow{2}{*}{MemConf2}&\textit{Mem}&1 channel, 2 DIMMs/channel, DDR4\creadyy{-2400}\\
&\textit{DRAM cache}&4GB stacked, 4-way, 4K pages, FBR\creadyy{~\cite{yu2017banshee}},  DDR4\creadyy{-2400}\\\hline
\multirow{2}{*}{OCM}&\textit{SERDES}&latency: 10/150/340 cycles \\
&\textit{Fiber}& latency: 30/60/90 cycles (2/4/6 meters roundtrip)\\\hline
NIC&\textit{40G PCIe~\creadyy{\cite{neugebauer2018understanding}}}& latency: 1050 cycles\\\hline
\end{tabular}
\label{table:zsim}
\end{table}

We calculate the SERDES \cready{link} latency values for the upcoming years. We estimate the \ruth{minimum} at 10 cycles, which assumes 3.2 ns serialization/deserialization latency~\cite{hmcpact}. We use 340 cycles (113ns) \ruth{maximum} latency reported in a previously demonstrated optical interconnection system \cite{proietti2018low}. We simulate rack distances of 2m, 4m, and 6m with a \ruthy{5 ns/m latency \cite{abali2015disaggregated}}, which {translates} into 30, 60, and 90 cycles latency in our system.

For the \joruge{40G NIC-based system configuration, we evaluate a scenario using a PCIe Network Interface Card (NIC) latency of 1050 cycles (350 ns) \cite{abali2015disaggregated} (a realistic NIC-through-PCIe latency is \cready{in} the order of microseconds \cite{neugebauer2018understanding}).} We evaluate the system-level performance of OCM  with applications \cready{from} four benchmark suites: (1)  SPEC06\creadyy{~\cite{henning2006spec}} using Pinpoints (warmup of 100 million instructions, and detailed region of 30 million instructions), (2) PARSEC\creadyy{~\cite{bienia2008parsec}} with \textit{native} inputs, (3) SPLASH2\creadyy{~\cite{bienia2008splash}} with \textit{simlarge} inputs, (4) SPEC17\creadyy{~\cite{bucek2018spec}} \emph{speed} with reference inputs, and (5) GAP graph benchmarks~\cite{BeamerAP15} executing 100 billion instructions with the \emph{Web} graph input, and 30 billion instructions with the \emph{Urand} graph input. The \emph{Urand} input has very poor locality between graph \cready{vertices} compared to the \emph{Web} input. \joruge{Table \ref{table:bench} lists the SPEC benchmark mixes we use in our \cready{multiprogrammed workload} evaluation.} \jorge{Table \ref{table:workload} summarizes the measured memory footprint values for all the benchmarks used in our evaluation.}

\begin{table}[h]
\scriptsize
\centering
\tabcolsep=0.07cm
\caption{Evaluated SPEC06 \cready{\&} SPEC17 benchmark mixes.}
{\color{black}\begin{tabular}{c|c|l}
\hline
\multirow{3.6}{*}{\rotatebox[origin=c]{90}{SPEC06}} & mix1 & \scriptsize soplex\_1, h264, gobmk\_3, milc, zeusm, bwaves, gcc\_1, omnetpp\\
&&\\[-2mm]
& mix2 & \scriptsize soplex\_1, milc,povray, gobmk\_2, gobmk\_3, bwaves, calculix, bzip2\_2 \\
&&\\[-2mm]
& mix3 & \scriptsize namd, gromacs, gamess\_1, mcf, lbm, h264\_2, hmmer, xalancbmk    \\
\hline
\multirow{3.6}{*}{\rotatebox[origin=c]{90}{SPEC17}} & mix1 & \scriptsize exchange2, cactus, gcc\_2, imagick, fotonik3d, xalancbmk, xz\_2, lbm\\
&&\\[-2mm]
& mix2 & \scriptsize gcc\_1, nab, lbm, leela, mcf, xz\_1, sroms, omnetpp\\
&&\\[-2mm]
& mix3 & \scriptsize xalancbmk, nab, cactus, mcf, imagick, xz\_1, fotonik3d, deepjeng\\
\hline
\end{tabular}}
\label{table:bench}
\end{table}

\begin{table}[h]
\scriptsize
\centering
\tabcolsep=0.07cm
\caption{Measured memory footprints.}
{\color{black}\begin{tabular}{r|l}
\hline
SPEC06~\cready{\cite{henning2006spec}} & \scriptsize \emph{MIX1}: 2.2 GB, \emph{MIX2}: 3.1 GB, \emph{MIX3}: 2.4 GB\\
\hline
SPEC17~\cready{\cite{bucek2018spec}}& \scriptsize \emph{MIX1}: 19.9 GB, \emph{MIX2}: 36.4 GB, \emph{MIX3}: 34.7 GB.\\
\hline
PARSEC~\cready{\cite{bienia2008parsec}} & \vtop{\hbox{\strut \scriptsize \emph{canneal}: 716.7 MB, \emph{streamcluster}: 112.5 MB, \emph{ferret}: 91.9 MB,}\hbox{\strut \scriptsize \emph{raytrace}: 1.3 GB, \emph{fluidanimate}: 672 MB}}\\
\hline
SPLASH~\cready{\cite{bienia2008splash}}& \vtop{\hbox{\strut \scriptsize \emph{radix}: 1.1 GB, \emph{fft}: 768.8 MB, \emph{cholesky}: 44.2 MB,}\hbox{\strut \scriptsize \emph{ocean\_ncp}: 26.9 GB, \emph{ocean\_cp}: 891.8 MB.}}\\
\hline
GAP~\cready{\cite{BeamerAP15}} & \scriptsize \emph{Urand} graph: 18 GB, \emph{Web} graph: 15.5 GB\\
\hline
\end{tabular}}
\label{table:workload}
\end{table}

\subsection{\lon{System-level Evaluation}} \label{sec:sysleveleval}

\noindent\textbf{\joruge{Multiprogrammed Evaluation.}} Figure \ref{fig:performance_single} shows the slowdown of OCM and \joruge{40G \cready{NIC-based}} disaggregated memory systems with MemConf1, compared to \joruge{a non-disaggregated MemConf1 baseline}, for three  mixes of SPEC06 benchmarks (Table \ref{table:bench}). Notice that a system with disaggregated main memory is expected to perform worse than the non-disaggregated baseline, because of the extra latency introduced by the interconnects (see Eq. \ref{time_total}).


We make two observations. First, the 40G NIC-based system is significantly slower than {our} OCM system, even though the Ethernet configuration \cready{we evaluate} is very optimistic (\cready{350 ns average latency, equivalent to 1050 cycles in Table \ref{table:zsim}}). OCM is up to $5.5\times$ faster than 40G NIC for the \ruth{minimum} SERDES latency, and \onur{$2.16\times$} faster for the \ruth{maximum} SERDES latency.
Second, the results show the feasibility of \cready{low-latency} disaggregation with OCM as future SERDES optimizations become available. OCM has an average slowdown (across all rack-distances) of \cready{only} $1.07\times$ compared to the baseline with a SERDES latency of 10 cycles, and $1.78\times$ average slowdown with a SERDES latency of 340 cycles.

\begin{figure*}[h]
  \centering
  \includegraphics[width=0.9\linewidth]{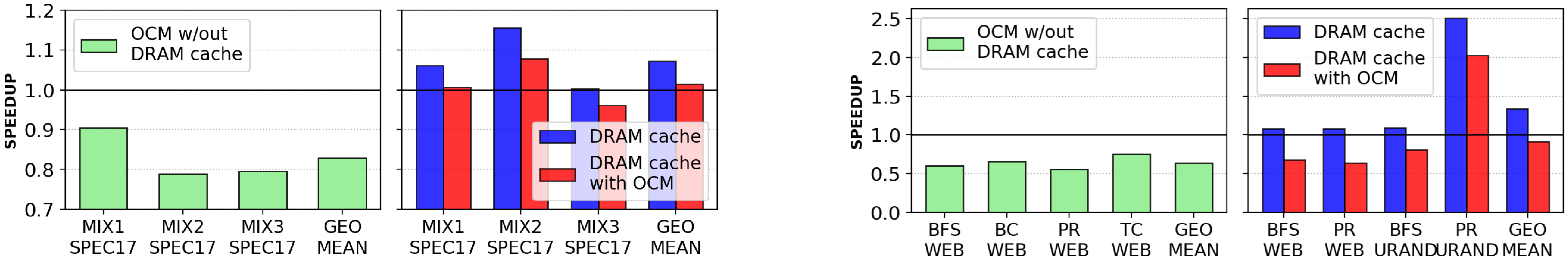}
  \caption{OCM \cready{speedup} results with 4m distance and a SERDES latency of 150 cycles (higher is better)\creadyy{, compared to a disaggregated baseline, with or without a DRAM cache}. Left: Speedup \cready{for} SPEC17. Right: Speedup \cready{for} \cready{GAP~\cite{BeamerAP15}} graph benchmarks}.
  \label{fig:graph_bench}
\end{figure*}

Figure~\ref{fig:graph_bench} shows the speedup of a disaggregated OCM  system (green bars) compared to a non-disaggregated baseline, both configured with MemConf1. Figure~\ref{fig:graph_bench} also shows the speedup of OCM with MemConf2 (red bars), and the speedup of a non-disaggregated system with MemConf2 (blue bars), both compared to a MemConf2 baseline without \cready{a} DRAM cache and without disaggregation. OCM has a conservative SERDES latency of 150 cycles, and a distance of 4m.

Figure~\ref{fig:graph_bench} (left) shows the results for SPEC17 mixes (see Table~\ref{table:bench}). We make two observations. First, the average slowdown of OCM without DRAM cache (green bars) is 17\%, which is in the same order as the SPEC06 results (Figure~\ref{fig:performance_single}). Second, \cready{with a DRAM cache}, the performance of the OCM disaggregated system (red bars), and the non-disaggregated system (blue bars) is very close, as the memory intensity of these benchmarks is not very high. As expected, the \cready{performance} of the disaggregated system is always lower than the non-disaggregated system.

\noindent\textbf{\joruge{Multithreaded Evaluation.}} Figure~\ref{fig:graph_bench} (right) shows the results for multithreaded graph applications. We make two observations. First, the maximum slowdown of OCM \cready{without a DRAM cache} (green bars) is up to 45\% (\emph{pagerank} (\emph{PR})), which is in the same order as SPEC17 results, despite the \emph{Web} input having very high locality. The extra latency of the OCM disaggregated system has a \cready{clear} negative effect on performance. Second, graph workloads dramatically benefit from using a DRAM cache (red and blue bars), e.g., \cready{\emph{PR} with \emph{Urand} input shows a speedup of $2.5\times$ compared to the baseline, which is 50\% lower speedup than the non-disaggregated scenario.} \cready{We believe that} the performance degradation of OCM with DRAM cache is still reasonable. However, adding a DRAM cache also brings new challenges that need further investigation \cready{in a disaggregated setting}, such as page replacement mechanisms \cready{and caching granularity\creadyy{~\cite{yu2017banshee,Li2017,Meza2012,yoon2014efficient,meza2013case,Yoon2012,Ramos2011,Jiang2010}}}.

\begin{figure*}[h]
  \centering
  \includegraphics[width=0.9\linewidth]{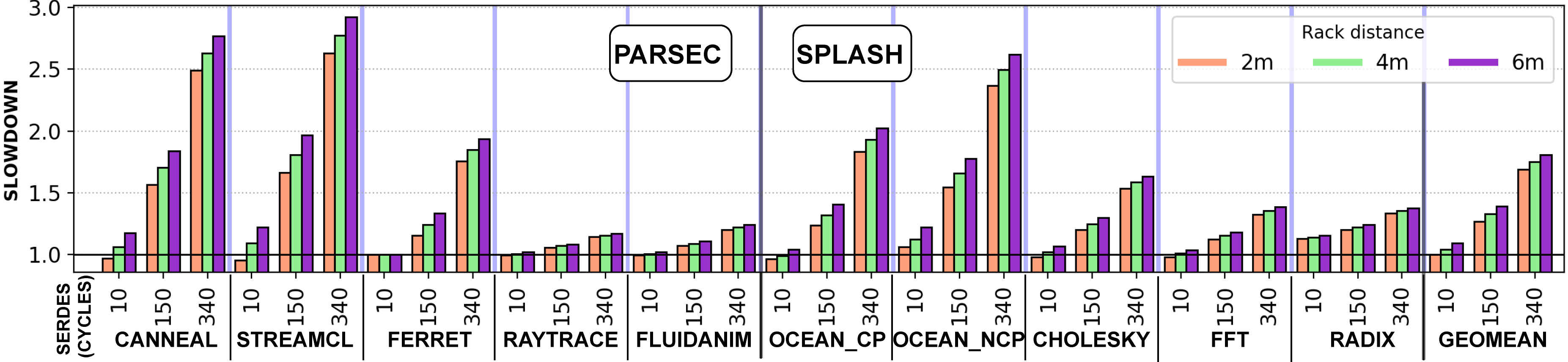}
  \caption{OCM slowdown compared to the baseline for PARSEC and SPLASH2 benchmarks (lower is better).}
  \label{fig:performance_multi}
\end{figure*}

Figure \ref{fig:performance_multi} shows the slowdown of OCM compared to \onur{the baseline}, using \joruge{MemConf1} with PARSEC and SPLASH2 benchmarks. \onur{We show results for the \onur{memory-bound} benchmarks only. We also test other compute-bound benchmarks (not shown in the figure) that show less than 5\% slowdown.} We make three observations. First, with the lower bound SERDES latency (10 cycles) and lowest rack distance (2 m), applications such as \cready{\textit{streamcluster}}, \textit{canneal} and \textit{cholesky}, {experience} an average 3$\%$ speedup. This \creadyy{small} improvement occurs as a result of $T_{mem}$ reduction ($tBL$ related) \cready{due to} splitting \cready{of a} cache line into two DIMMs. Second, the slowdowns increase slightly as distance increases. Third, with large rack-distance and \ruth{maximum} SERDES latency, the slowdown is significant. The highest slowdown measured is $2.97\times$ for \textit{streamcluster} at 6m and 340 SERDES cycles; the average slowdown is $1.3\times$ for SPLASH2 and $1.4\times$ for PARSEC.

\ruth{We conclude that OCM is very promising because of its \cready{reasonably} low latency overhead \cready{(especially with the use of a DRAM cache)}, and the flexibility of placing memory modules at large distances with small slowdowns.}

\subsection{SiP Link Evaluation}\label{sec:sip_section}

\cready{\creadyy{We evaluate the energy and area consumption of the SiP link} to allow the system designer to \creadyy{make tradeoffs about the use of} SiP devices in the computing system.}
\cready{We \creadyy{consider} unidirectional SiP links using PhoenixSim~\cready{\cite{rumley2016phoenixsim}} \cready{using} the parameters \creadyy{shown} in Table \ref{table:sip}. We estimate the minimum energy-per-bit consumption and the required number of MRRs for our model, given an aggregated optical bandwidth equivalent to the bandwidth required by \creadyy{DDR4-2400} DRAM memory.}

\begin{figure*}[h]
  \centering
    \includegraphics[width=0.9\linewidth]{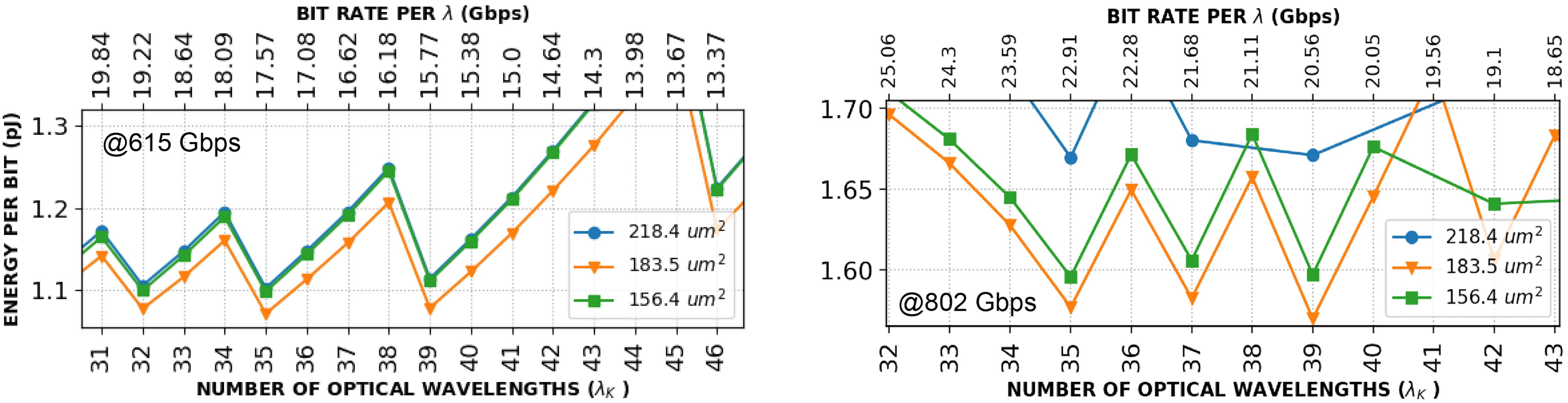}
    \caption{{SiP link energy-per-bit. Left: \creadyy{at} 615 Gbps \creadyy{bandwidth}, Right: \creadyy{at} {802} Gbps \creadyy{bandwidth}.} }
  \label{fig:optical_channels}
\end{figure*}

\ruth{A single DDR-2400 module requires 153.7 Gbps \creadyy{bandwidth}\creadyy{~\cite{jedecddr4}}. \cready{4} memory channels, with 2 DIMMs per channel in lockstep, require \creadyy{$\sim$}615 Gbps/link}. 
\cready{\creadyy{OCM's} maximum feasible  bandwidth (\creadyy{while remaining} CMOS compatible) is 802 Gbps using the parameters in Table \ref{table:sip}.}
More advanced modulation formats, such as PAM4~\creadyy{\cite{extremescale}}, can be used to achieve higher aggregated bandwidth. Figure \ref{fig:optical_channels} shows the energy-per-bit \cready{results} (y-axis), and the aggregated bandwidth. The  aggregated \cready{link} bandwidth is the multiplication of the number of $\lambda$ (bottom x-axis values), and \ruth{the aggregated bitrate (top x-axis values), i.e., a higher number of $\lambda s$ implies a lower bitrate per $\lambda$}. We consider \cready{three feasible \creadyy{and efficient} MRR sizes}
in our model: 156.4 (\jorrge{green}), 183.5 (orange), and $218.4\ \mu m^2$ (\jorrge{blue}).

\onur{In OCM with 615 Gbps links, the minimum energy consumption overhead \cready{compared to the electrical memory system} is 1.07 pJ/bit for 35 optical wavelengths ($\lambda$) per link, each $\lambda$ operating at 17.57 Gbps. In OCM with {802} Gbps links, the minimum energy consumption is 1.57 pJ/bit for 39 $\lambda$s per link, each $\lambda$ operating at 20.56 Gbps.}

\jorge{We make three observations from Figure \ref{fig:optical_channels}. 
First, as in electrical systems, it is expected that a higher bandwidth per link increases the link energy-per-bit consumption. However, the optical energy\cready{-per-bit} is lower compared to electrical systems. For reference, the energy-per-bit of a DDR4-2667 DRAM module is 39 pJ \cite{ddr4energy}; thus, the energy\cready{-per-bit} caused by \cready{an additional} SiP link in the memory subsystem is less than 5\%. Second, there is a non-smooth behavior on the energy-per-bit curves due to the energy consumption model of the optical receiver, which depends on the data rate. In our model, we set the photodetector current to a minimum value. \cready{As} the data rate increases, the received signal \creadyy{becomes less distinguishable from noise}. Our model forces the photocurrent to step into a new minimum value \cready{to avoid this,} causing the repeated decrease and increase of the energy-per-bit values~\cite{bahadori2017energy}.}
For both SiP links, the $183.5\ \mu m^2$ rings consume the lowest energy. The estimated area overhead is 51.4E-3 mm$^2$ with $2\times 615$ Gbps links, and 57.3E-3 mm$^2$ with $2\times802$ Gbps links.
In our case study of 4 DDR4 memory channels, OCM uses \cready{fewer} physical interconnects (optical fibers) than 40G PCIe NIC links (copper cables). \cready{In other words}, to achieve the required aggregated \creadyy{link} bandwidth, we require 2 optical fibers with OCM or 30 copper cables with 40G PCIe NICs.

\cready{We conclude that a bidirectional SiP link, formed by two unidirectional links \creadyy{using} current SiP devices, can fit the bandwidth requirements of commodity \creadyy{DDR4} DRAM modules. OCM incurs a low energy overhead of \creadyy{only} 10.7\% compared to \creadyy{a non-disaggregated DDR4} DRAM memory (the energy consumption of current \creadyy{DDR4} DRAM technology is $\sim10$pJ/bit \cite{extremescale}).}


\section{Related Work}\label{sec:relatedwork}

To our knowledge, \cready{this is the first work to} propose an optical point-to-point disaggregated main memory system for modern DDR memories that (1) evaluates a SiP link with state-of-the-art optical devices, (2) \creadyy{demonstrates that OCM incurs} \creadyy{only 10.7\% energy overhead compared to a non-disaggregated DDR4 DRAM memory}, and (3) \cready{quantifies} the performance implications of the proposed optical links at \cready{the system level on commonly-used} application workloads.


\jorge{Brunina et al. \cite{brunina1,brunina3} introduce the concept of optically connected memory in a mesh topology connected with optical switches. Both works propose point-to-point direct access to the DRAM modules using Mach Zender modulators. \cready{These} works motivate our study in optically connected memory. 
\cready{Brunina et al. \creadyy{\cite{brunina4}} also} experimentally demonstrate that microring modulators can be used for optically connecting DDR2 memory. Our work \cready{builds} on \cite{brunina4} to \cready{design} the \cready{microring} modulators used in our SiP links. \cready{There are several recent works \cite{bahadori2016comprehensive,extremescale,bahadori2017energy} that propose analytical models of the microring used in our SiP links}.
\cready{Anderson et al.~\cite{anderson2018reconfigurable} extend the work of Brunina et al. \cite{brunina1,brunina3,brunina4} to experimentally demonstrate the optical switches using FPGAs for accessing memory.}
}





\jorge{\creadyy{These prior} works \cite{anderson2018reconfigurable,brunina4,brunina3,brunina1} are \creadyy{all} experimental demonstrations to show photonic capabilities. In contrast, our work \cready{addresses} three important questions \cready{prior work does not}: (1) How many optical devices (i.e., MRRs) do we need for current DDR technology? \cready{(Section \ref{sec:sip_section})}, (2) What is the energy and area impact on the system? \cready{(Section \ref{sec:sip_section})}, and (3) How \cready{does} the processor \cready{interact} with a disaggregated memory subsystem? \cready{(Section \ref{sec:sysleveleval})}.}


\jorge{\cready{Some other} works, such as \cite{weiss2014optical,zhu2019flexible}, point out, \cready{without \creadyy{evaluation}}, that \creadyy{existing} disaggregation protocols (i.e., PCIe and Ethernet) could lead to high-performance loss. \cready{Our work uses} system-level simulation to measure the performance overhead \creadyy{of such protocols.} \creadyy{We} propose to alleviate the optical serialization overhead by using \creadyy{the DDR protocol} (Section~\ref{sec:archoverview}).}
As photonic integration improves, we \cready{believe} that the optical point-to-point links \cready{will become} the main candidate for interconnecting disaggregated memory. With our PhoenixSim \cite{rumley2016phoenixsim} model, we explore the design of SiP links based on DDR requirements. Our proposal can be used to improve existing PCIe+photonics works, such as \cite{yan2016all}.


\cready{\jorge{\cready{Yan et al.~\cite{yan2016all} propose} a PCIe Switch and Interface Card (SIC) to replace Network Interface Cards (NIC) for disaggregation. SIC is composed of commercial optical devices and is capable of interconnecting server blades in disaggregated data centers. \cready{The evaluated} SIC shows a total roundtrip latency up to 426 ns. \cready{In contrast}, the scope of our work is point-to-point \cready{DDR DRAM} disaggregation without PCIe or other additional protocols.}}

Other \cready{related} prior works (1) explore silicon photonics integration with a many-core chip in an optical network-on-chip design~\cite{batten2009building}, (2) {propose the design of a DRAM chip with photonic inter-bank communication~\cite{beamer2010re}}, (3) {present an optoelectronic chip for communication in disaggregated systems with 4-$\lambda$ and an energy consumption of 3.4 pJ/bit~\cite{akhter2017wavelight}}, (4) evaluate \cready{a} memory disaggregation architecture with optical switches focusing on re-allocation mechanisms \cite{Zervas:18}, (5) analyze \cready{the} cost viability of optical memory disaggregation~\cite{abali2015disaggregated}, and (6) evaluate memory disaggregation using software mechanisms with high latency penalties in the order of $\mu$s \cite{gu2017efficient}. \cready{Unlike \creadyy{\cite{Zervas:18,akhter2017wavelight,beamer2010re,abali2015disaggregated,gu2017efficient}}, our work evaluates i) system performance with real applications, ii) \creadyy{the} design of the SiP link for DDR DRAM requirements, and iii) SiP link energy for a disaggregated memory system.}

\section{Conclusions}

\jorge{\cready{We} propose and evaluate \cready{Optically Connected Memory (OCM)}, a new optical architecture for \cready{disaggregated main} memory systems, compatible with current \cready{DDR} DRAM technology. OCM uses a \cready{Silicon Photonics (SiP)} platform that enables memory disaggregation with low energy-per-bit overhead. 
\cready{Our evaluation shows that, for the bandwidth required by current DDR standards, OCM has significantly better energy efficiency than conventional electrical NIC-based communication systems, and it  incurs  a  low  energy  overhead  of \creadyy{only} 10.7\%  compared  to DDR  DRAM  memory.
}
\cready{Using} system-level \cready{simulation} to evaluate our OCM model \cready{on real applications}, \cready{we find that OCM performs} 5.5 times faster than a 40G NIC-based disaggregated memory.} 
\cready{We  conclude that OCM is a promising step towards future data centers with disaggregated main memory.}

\bibliographystyle{IEEEtranS}

\bibliography{bibliography}




\end{document}